\documentclass[12pt]{article}
\usepackage[latin1]{inputenc}
\usepackage[all]{xy}
\usepackage{graphicx}
\usepackage{latexsym}
\usepackage{color}
\usepackage{epsfig}
\usepackage{amsmath}
\usepackage{amsfonts}
\usepackage{amssymb}
\usepackage{amsthm}
\usepackage{mathabx}
\usepackage{MnSymbol}
\usepackage[nospace]{cite}

\newcommand{\transi}{{ \cap \!\!\!\!\!\! - \!\!\!\!-}}

\def\l[{\left[}
\def\r]{\right]}

\def\ba{\begin{array}}
\def\ea{\end{array}}
\def\beq{\begin{equation}}
\def\eeq{\end{equation}}
\def\bea{\begin{eqnarray}}
\def\eea{\end{eqnarray}}

\begin{document}

\pagestyle{empty}
\setcounter{page}{0}

\vspace{2cm}

\begin{center}
{\Large {\bf Topological gauge fixing II}}

\vspace{0.3cm}

{\large {\bf A homotopy formulation}}
\\[1.5cm]

{\large L.~Gallot, E.~Pilon and F.~Thuillier}

\end{center}

\vskip 0.7 truecm

{\it LAPTH, Universit\'e de Savoie, CNRS, 9, Chemin de Bellevue, BP 110, F-74941 Annecy-le-Vieux cedex, France}.

\vspace{2cm}

\begin{abstract}
We revisit the implementation of the metric-independent Fock-Schwinger gauge in the abelian Chern-Simons field theory defined in ${\mathbb{R}}^3$ by means of a homotopy condition. This leads to the lagrangian $F \wedge hF$ in terms of curvatures $F$ and of the Poincar\'e homotopy operator $h$. The corresponding field theory provides the same link invariants as the abelian Chern-Simons theory. Incidentally the part of the gauge field propagator which yields the link invariants of the Chern-Simons theory in the Fock-Schwinger gauge is recovered without any computation.
\end{abstract}

\newpage
\pagestyle{plain} \renewcommand{\thefootnote}{\arabic{footnote}}

Linking numbers are related to expectation values of Wilson loops in the abelian Chern-Simons theory \cite{G,GT}. The computation of these expectation values involves the propagator of the gauge field which in turn requires a gauge fixing. In the covariant gauge \cite{GPT}, the gauge field correlator is nothing but the Gauss linking density of the linking number. In a companion article \cite{GGPT} the Fock-Schwinger (aka radial) gauge $x^\mu A_\mu(x) = 0$ was considered. This gauge fixing is "topological" in the sense that it is metric independent. Here we would like to present an alternative approach using the curvature $F_A = dA$ and its correlator instead of the gauge potential $A$. This is achieved by considering the Poincar\'e Homotopy gauge condition $hA = 0$ which is equivalent to the Fock-Schwinger gauge condition.

\vspace{3mm}

\noindent The Poincar\'e homotopy $h: \Omega^p \rightarrow \Omega^{p-1}$ ($p > 0$) in $ \mathbb{R}^n$ is the operator defined by \cite{C-B}:
\begin{eqnarray}
(h \omega)(x)
= \frac{1}{(p-1)!}\left( \int_{0}^{1} dt \, t^{p-1} \, x^{\nu} \, \omega_{\nu \mu_2 \cdots \mu_p}(tx) \right) \, dx^{\mu_2} \wedge \cdots \wedge dx^{\mu_p} \, ,
\label{egt1}
\end{eqnarray}
where $\Omega^p $ denotes the space of $p$-forms on $ \mathbb{R}^n$. It satisfies the fundamental identity:
\begin{equation}
dh + hd = 1.
\label{egt2}
\end{equation}
Since the space ${\cal A}^\infty$ of smooth $U(1)$ gauge fields in ${\mathbb{R}}^3$ identifies with $\Omega^1$ the {\it Poincar\'e Homotopy gauge} in ${\cal A}^\infty$ is defined by:
\begin{eqnarray}
hA = 0 \, .
\label{egt3}
\end{eqnarray}
This yields a subspace ${\cal A}^\infty_h$ of ${\cal A}^\infty$. In spherical coordinates $x = r \hat{r}$ the Fock-Schwinger condition reads $A_r(x) = 0$ whereas the Poincar\'e condition (\ref{egt3}) may be rewritten
\bea
\int_0^r ds \; A_r(s \, \hat{r}) = 0
\label{egt4}
\eea
The Fock-Schwinger condition implies that $A_r(s \, \hat{r}) = 0$ for any $s \neq 0$, hence condition (\ref{egt4}). Conversely the derivative of (\ref{egt4}) with respect to $r$ readily leads to the Fock-Schwinger condition. This proves the equivalence of the Fock-Schwinger and Poincar\'e Homotopy gauges.

\noindent Due to (\ref{egt2}), for any $A \in {\cal A}^\infty_h$ one has:
\begin{equation}
F_A := dA = (dh + hd)F_A = dhF_A \, ,
\label{egt5}
\end{equation}
since $dF_A = d^2 A = 0$. The space ${\cal F}^\infty$ of smooth $U(1)$ curvatures $F_A$ in in ${\mathbb{R}}^3$ identifies with $\Omega^2_0$, the space of closed $2$-forms in ${\mathbb{R}}^3$. In ${\cal A}^\infty_{h}$ one has:
\begin{equation}
A = (dh + hd)A = hdA = hF_A \, .
\label{egt6}
\end{equation}
Equations (\ref{egt5}) and (\ref{egt6}) imply that ${\cal F}^\infty \simeq_h {\cal A}^\infty_h$ with $h = d^{-1}$ on ${\cal F}^\infty$. This is nothing but Poincar\'e lemma. In Quantum Field Theory fields are not smooth but rather distributions, more precisely de Rham currents \cite{dR}. We denote by ${\cal A}_h$ and ${\cal F}$ the spaces of singular $U(1)$ gauge fields and curvatures. The de Rham derivative $d$ and the Poincar\'e homotopy operator $h$ both extend to currents, and so does Poincar\'e's lemma \cite{dR}, so that $h = d^{-1}$ still holds on ${\cal F}$.

\vspace{3mm}

\noindent The gauge fixed Chern-Simons action takes the form:
\begin{equation}
{\cal S}_{CSh} = {\cal S}_{CS} +{\cal S}_{GF} = 2 \pi k \, \left\{ \int_{{\mathbb{R}}^3} A \wedge d A + \int_{{\mathbb{R}}^3} B \wedge \, ^\star hA\, \right\} ,
\label{egt7}
\end{equation}
where $^\star$ denotes the Euclidean Hodge star operator. In principle the action (\ref{egt7}) also contains a ghost term \cite{GGPT}. As the ghosts do not couple to the gauge field they may be integrated out explicitly amounting to an overall normalization. We omit them here for the sake of simplicity.

\noindent The generating functional of the $U(1)$ Chern-Simons theory is given by:
\begin{equation}
{\cal Z}_{CSh} (j)
=
\int {\cal D} A \, {\cal D} B \,
 \; e^{i {\cal S}_{CSh} \, + \, 2 i \pi \int A \wedge j}
\label{egt8}
\end{equation}
where the source $j$ for the gauge field $A$ is a (smooth) $2$-form. In order to reformulate the $U(1)$ Chern-Simons theory in the Poincar\'e Homotopy gauge as a theory involving curvatures instead of gauge potentials let us insert
\begin{equation}
1 =  \int {\cal D} F \, \delta \left( F - dA \right)
\label{egt9}
\end{equation}
into the generating functional ${\cal Z}_{CSh}$, the functional integral in (\ref{egt9}) being performed on the space ${\cal F}$. The constraint $\delta \left( F - dA \right)$, originally set on $F$, can be translated into a constraint on $A$ by writing
\begin{eqnarray}
\delta \left( F - dA \right) = \delta \left( d ( A - hF ) \right)
= \Xi^{-1} \, . \, \delta \left( A - hF \right) \,
\label{egt10}
\end{eqnarray}
where $\Xi$ denotes the determinant of the restriction of $d$ to ${\cal A}_{h}$. Using equation (\ref{egt6}) and $hh F =0$, the action ${\cal S}_{CSh}$ can be recasted into:
\begin{equation}
{\cal S}_{F} = 2 \pi k \left\{ \int hF \wedge F + \int B \wedge \, ^\star h h F \right\} = 2 \pi k  \int hF \wedge F \, .
\label{egt11}
\end{equation}

\noindent If a source $j$ of $A$ is closed, \textit{i.e.} such that $dj = 0$, then according to Poincar\'e's lemma $j = d \psi$ for some $1$-form $\psi$, and therefore:
\begin{equation}
\int A \wedge j = \int A \wedge d\psi = \int dA \wedge \psi = \int F \wedge \psi  \, .
\label{egt12}
\end{equation}
Note that the closeness of $j$ ensures the gauge invariance of $e^{2 i \pi \int A \wedge j}$.

\noindent The restriction of ${\cal Z}_{CSh}$ to closed sources then reads:
\begin{eqnarray}
{\cal Z}_{CSh} (j)
& = &
\int {\cal D} F \, {\cal D} A \, {\cal D} B \,
 \; \, \Xi^{-1} \, \delta (A - hF) \,
e^{  i {\cal S}_{hF}  } \nonumber \\
& = & ( \int {\cal D} A {\cal D} B \, \Xi^{-1} ) \, \int {\cal D} F \, e^{  i {\cal S}_{hF}  \, + \, 2i \pi \int F \wedge \psi } \, .
\label{egt13}
\end{eqnarray}
The functional integral over $A$ and $B$ gives rise to an overall normalization factor, whereas the remaining factor is the generating functional for the field theory with action ${\cal S}_{F}$:
\begin{eqnarray}
{\cal Z}_F (\psi)
& = &
\int {\cal D} F \,
e^{ i {\cal S}_{hF}  \, + \, 2i \pi \int F \wedge \psi} \, .
\label{egt14}
\end{eqnarray}
The $1$-form $\psi$ is a source of $F$. Note that this generating functional satisfies:
\bea
{\cal Z}_F (\psi + d \lambda) = {\cal Z}_F (\psi) \,
\label{egt15}
\eea
for any $0$-form, {\it i.e.} function,  $\lambda$. This symmetry of ${\cal Z}_F$ reminds of the gauge invariance of the original Chern-Simons theory.

\vspace{3mm}

\noindent To generate an invariant of a link $L$ in $\mathbb{R}^3$, one considers the expectation value of its holonomies:
\begin{equation}
\left< {\cal W}(L) \right>_{CSh}
= \frac{1}{{\cal N}_{CSh}}
\int {\cal D} A \, {\cal D} B \,
 \; e^{i {\cal S}_{CSh} \, + \, 2 i \pi \int_{L} A }
\label{egt16}
\end{equation}
with ${\cal N}_{CSh} = {\cal Z}_{CSh} (0)$. Yet a knot $C$ in ${\mathbb{R}}^3$ canonically defines a closed de Rham $2$-current $J_C$, {\it i.e.} a closed $2$-form with distributional coefficients \cite{dR}, in such a way that $e^{2 i \pi \int_{C} A } = e^{2 i \pi \int A \wedge J_C }$. As for sources, the closeness of $J_C$, or equivalently of $C$, ensures the gauge invariance of ${\cal W}(L)$. Since Poincar\'e's lemma also holds for currents, we have:
\begin{equation}
\int A \wedge J = \int A \wedge d\Psi = \int dA \wedge \Psi = \int F_A \wedge \Psi  \, ,
\label{egt17}
\end{equation}
for some $1$-current $\Psi$. Furthermore if two $1$-currents $\Psi$ and $\Psi'$ satisfy $d\Psi = J_C = d\Psi'$ then $\Psi' = \Psi + d \Lambda$ for some $0$-current $\Lambda$. This reproduces at the level of currents the geometrical property that any knot in ${\mathbb{R}}^3$ is bounding a surface, and if two surfaces in ${\mathbb{R}}^3$ share the same boundary their difference encloses a volume.

\noindent Equation (\ref{egt17}) suggests to replace (\ref{egt16})  by:
\begin{equation}
\left< {\Phi}(\Sigma) \right>_F
= {\cal N}_{F}
\int {\cal D} F \,
 \; e^{i {\cal S}_{F} \, + \, 2 i \pi \int_{\Sigma} F } = e^{2 i \pi k  \int hF \wedge F \, + \, 2 i \pi \int F \wedge \Psi_{\Sigma} }
\label{egt18}
\end{equation}
where $\Sigma$ is a surface in $\mathbb{R}^3$, $\Psi_{\Sigma}$ its de Rham $1$-current, and ${\cal N}_{F} = {\cal Z}_{F} (0)$. As for sources of $F$, we can identify $\Psi$ and $\Psi + d \Lambda$ since two such $1$-currents generate the same "quantum flux" $e^{2 i \pi \int F \wedge \Psi}$. Quantum fluxes are thus defined on ${\cal J}^1 \equiv \Omega'^1 / d \Omega'^0$ rather than on $ \Omega'^1 $, with $\Omega'^p$ denoting the space of $p$-currents in $\mathbb{R}^3$.
Let us point out the similarity between ${\cal J}^1$ and ${\cal A}_h$, each element of the latter being a particular representative of an element of the former. From now on $\Psi$ will indistinctly denote a class in ${\cal J}^1$ or a representative $1$-current of this class.

\noindent Thanks to the quadratic form of the action ${\cal S}_{F} $, the functional integral (\ref{egt18}) can be computed explicitly giving:
\begin{eqnarray}
\left< {\Phi}(\Sigma) \right>_F = \exp \left\{\frac{(2 i \pi)^2}{2} \int \Psi_{\Sigma}(x) \wedge \left( \left< F(x) \, F(y) \right> ^\star \Psi_{\Sigma}(y) \right) \right\} \, ,
\label{egt19}
\end{eqnarray}
where the curvature propagator is
\begin{eqnarray}
\left< F(x) \, F(y) \right> ^\star & = & \frac{i}{4 \pi k} \, h^{-1}_{y} \, \delta^{(3)}(y-x) = \frac{i}{4 \pi k} \, d_{y} \, \delta^{(3)}(y-x) \, ,
\label{egt20}
\end{eqnarray}
since equation (\ref{egt2}) implies that $h^{-1}_{y} = d_y$ on ${\cal J}^1$. Consequently if $L$ is a link and $\Sigma$ is a surface bounded by $L$ then:
\begin{eqnarray}
\left< \Phi (\Sigma) \right>_F
& = &
\exp \left\{ - \frac{2 i \pi}{4 k} \int \Psi_{\Sigma} \wedge h^{-1} \Psi_{\Sigma} \right\} \, . \label{egt21}
\end{eqnarray}
Since $h^{-1} \Psi_\Sigma = d \Psi_\Sigma = J_L$, one has:
\begin{eqnarray}
\Psi_{\Sigma} \wedge h^{-1} \Psi_{\Sigma} =  \Psi_{\Sigma} \wedge d \Psi_{\Sigma} \, .
\label{egt22}
\end{eqnarray}
Thus:
\begin{eqnarray}
\int \Psi_{\Sigma} \wedge h^{-1} \Psi_{\Sigma} =  \Sigma \; \transi \, L \, ,
\label{egt23}
\end{eqnarray}
where $\transi$ denotes the transverse intersection of a surface and a curve in $\mathbb{R}^3$. Once a framing of $L$ (or rather of its component knots) is given, intersection (\ref{egt23}) is nothing but the linking of $L$ with itself. The latter is also the expectation value of the Wilson loop of $L$ in the CS theory \cite{GT}, \textit{cf.} (\ref{egt16}):
\begin{eqnarray}
\left< \Phi (\Sigma) \right>_F = \exp \left\{ - \frac{2 i \pi}{4 k} \; lk(L,L) \right\} = \left< {\cal W}(L)  \right> _{CSh} \label{egt24} \, .
\end{eqnarray}
The first equality of (\ref{egt24}) is obtained using the theory defined by ${\cal S}_F$, whereas the last one comes from the original $U(1)$ Chern-Simons theory in the Poincar\'e Homotopy gauge \cite{GGPT}. This set of equations establishes the equivalence of the two theories at the level of the observables considered: "quantum fluxes" for ${\cal S}_F$ and holonomies for ${\cal S}_{CSh}$.

\vspace{3mm}

\noindent As byproduct, the propagator $\left< A(x) \, A(y) \right>$ for the Chern-Simons theory in the Poincar\'e Homotopy gauge \cite{GGPT} can be obtained from (\ref{egt20}) by simply writing:
\begin{eqnarray}
\left< A(x) \, A(y) \right>^\star = \left< h_x F(x) \, h_y F(y) \right>^\star = \frac{i}{4 \pi k} \, \delta(y-x) \, h_{x} \, ,
\label{egt25}
\end{eqnarray}
which coincides with the propagator computed in \cite{GGPT}.

\vspace{3mm}

\noindent All we have presented here extends to the $U(1)$ Chern-Simons theory in ${\mathbb{R}}^{4n+3}$ introduced in \cite{GPT}.

\end{document}